\documentclass[conference]{IEEEtran}
\IEEEoverridecommandlockouts
\usepackage[colorlinks = true, linkcolor = blue, urlcolor = blue, citecolor = blue, anchorcolor = blue]{hyperref}
\usepackage{amsmath,amssymb,amsfonts}
\usepackage[color=green]{todonotes}
\usepackage{algorithmic}
\usepackage{graphicx}
\usepackage{textcomp}
\usepackage{xcolor}
\usepackage{balance}
\usepackage[figurename=Figure, center]{caption}
\usepackage{subcaption}

\def\BibTeX{{\rm B\kern-.05em{\sc i\kern-.025em b}\kern-.08em
    T\kern-.1667em\lower.7ex\hbox{E}\kern-.125emX}}
    
\usepackage{amsthm}
\theoremstyle{definition}
\newtheorem{defn}{Definition} 

\usepackage{verbatim}
\usepackage{fancyvrb}
\usepackage{comment}    
\begin{document}

\title{Metamorphic Testing in Autonomous System Simulations\\
\thanks{This work was partially supported by the Austrian Science
Fund (FWF), under grant I 4701-N} 
}

\author{\IEEEauthorblockN{Jubril Gbolahan Adigun}
\IEEEauthorblockA{\textit{Department of Computer Science} \\
\textit{University of Innsbruck}\\
Innsbruck, Austria \\
jubril.adigun@uibk.ac.at}
\and
\IEEEauthorblockN{Linus Eisele}
\IEEEauthorblockA{
\textit{Department of Computer Science} \\
\textit{University of Innsbruck}\\
Innsbruck, Austria \\
eisele.linus@gmail.com}
\and
\IEEEauthorblockN{Michael Felderer}
\IEEEauthorblockA{\textit{Department of Computer Science} \\
\textit{University of Innsbruck}\\
Innsbruck, Austria  \\
michael.felderer@uibk.ac.at}

}

\maketitle



\begin{abstract}
Metamorphic testing has proven to be effective for test case generation and fault detection in many domains. It is a software testing strategy that uses certain relations between input-output pairs of a program, referred to as metamorphic relations. This approach is relevant in the autonomous systems domain since it helps in cases where the outcome of a given test input may be difficult to determine. In this paper therefore, we provide an overview of metamorphic testing as well as an implementation in the autonomous systems domain. We implement an obstacle detection and avoidance task in  autonomous drones utilising the GNC API alongside a simulation in Gazebo. Particularly, we describe properties and best practices that are crucial for the development of effective metamorphic relations. We also demonstrate two metamorphic relations for metamorphic testing of single and more than one drones, respectively. Our relations reveal several properties and some weak spots of both the implementation and the avoidance algorithm in the light of metamorphic testing. The results indicate that 
metamorphic testing has great potential in the autonomous systems domain and should be considered for quality assurance in this field.
\end{abstract}

\vspace{3mm}

\section{Introduction}
There is a growing adoption of autonomous systems (ASs) in our daily lives and this requires that software and system engineers ensure that ASs are adequately tested. One of the now frequently used methods for testing is simulation testing in which different approaches such as metamorphic testing (MT), fuzz testing, search-based testing, machine learning testing and model-based testing are conducted. In this work, we focus on the metamorphic testing technique as it has effectively been used across many domains \cite{survey, mainArticle}. Evidently, MT has been receiving an increasing amount of attention and has led to several successful implementations, but not as many applications in autonomous systems. Moreover, there appears to be a limited amount of studies when it comes to the identification of metamorphic relations (MRs) despite there being several open questions on how to construct effective MRs \cite{relationIdentification}. We believe that the lack of applications in autonomous systems is unjustified since MT may be a very appropriate strategy for testing autonomous systems. Consequently, the contributions of this work are;
\begin{itemize}
    \item We introduce MT in autonomous systems and define several MRs.
    \item We develop an obstacle avoidance task in an autonomous drone simulation.
    \item We implement the test cases to demonstrate the defined MRs.
    \item We provide detailed technical information for setting up the simulation used in this work.
    \item We provide evidence that MT is an adequate testing strategy in the simulation of ASs
\end{itemize}

The remaining part of this paper is organised as follows: Section \ref{sec:background} presents a theoretical background to some of the concepts mentioned in this study. The approach in terms of the testing technique is described in Section \ref{sec:preliminaries}. In Section \ref{sec:implementation}, we describe the implementation infrastructure as well as the tools used for the simulation and discuss the results in Section \ref{sec:results}. Further, we present some related work in metamorphic testing and highlight studies carried out using simulations in Section \ref{sec:related}. Finally, we conclude and present some ideas for future work in Section \ref{sec:conclusion}.

\section{Background}\label{sec:background}
Here, we provide background information about the concepts used throughout the paper.
\subsection{Software Testing}
Software testing is an essential practice widely applied during development in order to detect faults in a program. This usually consists of running the program under test on selected input data, producing corresponding output data which are then examined. The next step is to determine whether the program is correct or not. The mechanism which reveals this is known as the (test) oracle, and the common assumption that an oracle is trivially given is called the oracle assumption \cite{nontestablePrograms}. While this assumption is justified in many cases, there do exist programs that produce complex output such as compilers, simulators, or search engines. For these programs there may not exist an oracle or it may not be feasible to use, which we refer to as the oracle problem \cite{mainArticle}. Programs which suffer from the oracle problem are often referred to as ``untestable'' since one can hardly benefit from testing when it is impossible or too inefficient to classify a program’s output as (in)correct \cite{nontestablePrograms}. Autonomous systems are a critical domain in this regard since it can be immensely difficult to decide whether the behaviour of an autonomous system meets its specified requirements. Furthermore, autonomous drones are becoming increasingly popular for purposes like surveillance or search operations, while software testing in this field is complex \cite{AIdrones}.
\subsection{Simulation Testing}
Simulation testing allows designers, developers, engineers, and researchers to mimic real world systems and operational behaviours in a controlled virtual or hybrid environment. It has gained popularity in the scientific world and been applied severally in various complex cyber-physical systems (CPSs) such as in the autonomous vehicles \cite{10.1007/978-3-658-21194-3_82}, Internet of Things (IoT) and distributed systems domains. Contemporary testing methods such as rule-based techniques, semi-formal and formal verification often require trivialising systems and do not always provide adequate reliability necessary when testing complex systems \cite{9645356}. More so, these models operate at a rather abstract level depicting states of the system or some other timed automata without giving insight about the operational environment in which a system would act. In this case, simulations offer a more robust exploratory platform as physical effects and environmental details can be observed and analysed. They can be used for understanding functional requirements of a system as well as to analyse the safety concerns of the system in operation as done by Huck et al. \cite{9645356}. Simulations immensely reduce cost, time and risk for development and testing compared to a real-world environment and generally, an untested implementation should not be blindly applied to a real AS.
\subsection{Metamorphic Testing}
Metamorphic Testing (MT) is a testing technique that aims to alleviate the oracle problem by checking whether the inputs and outputs of a program across multiple executions satisfy certain relations, rather than directly investigating its input-output behaviour, as conventional testing strategies would do \cite{mainArticle}. These relations, basically properties of the system under test, are known as Metamorphic Relations (MRs) \cite{survey}.
For instance, suppose one were testing an e-commerce site and performs an initial search for a particular product across all regions where the business is located. It is not feasible to verify if the search result is fully correct, but one could use a follow up action to determine if the search result is problematic or not. So, one could note the set of items returned, say $X$, from the initial search. One may then apply another search filter e.g. a specific location filter to the search parameters such as ``Region: My Location'' and then run the search again. we expect that the returned set of products, say $Y$ should be a subset of the former result $X$ i.e. $Y \subset X$. This represents a simple example of a metamorphic relation where we may tell that something is wrong in our system if for instance, the cardinality of $Y$ were greater than the cardinality of $X$ even though we may not know exactly where the error may have come from.


\section{Preliminaries}\label{sec:preliminaries}
In this section, we define the relevant terminologies used in MT. We also describe the process of identification of metamorphic relations and (MRs) and briefly discuss how to construct effective MRs. 

\subsection{Terminology}
We combine the notations used by Chen et al. in \cite{founders} and \cite{challengesOpportunities}. The resulting notation has been slightly generalised and further elaborated in some aspects.

Let $i,\, n,\, m \in \mathbb{N} \,\wedge\, m > n > i$, and let $P$ be a program that implements an algorithm or function $f$. Given $x_i$ as input for $f$, the respective output is denoted as $f(x_i)$. 
The result of executing $P$ with test case $\overline{x_i}$ is denoted by $P(\overline{x_i})$. 
For a test case $\overline{x_i}$ with the result $P(\overline{x_i})$, a number of ``next'' test cases denoted as $\overline{x_{i+1}},\, \overline{x_{i+2}},\, \dots,\, \overline{x_n}$ may be constructed.   
A sequence of elements $x,\,y,\,z$ (i.e. test cases, inputs, outputs, or results) is denoted as $\langle x,\,y,\,z \rangle$. 

\begin{defn}[\emph{Metamorphic Relation}]
	\label{def:MR}
	{\ \\} 
	A metamorphic relation (MR) is a necessary property\footnote{A property that can be logically deduced from the algorithm \cite{challengesOpportunities}.} 
	of $f$ over a sequence of inputs $\langle x_i,\, x_{i+1},\, \dots,\, x_n\rangle$ and their corresponding outputs $\langle f(x_i),\, f(x_{i+1}),\, \dots,\, f(x_n)\rangle$.
	It can also be expressed as a relation $R$, where 
	$R(x_i,\, x_{i+1},\, \dots,\, x_n,\, f(x_i),\, f(x_{i+1}),\, \dots, f(x_n)) 
	\\\Rightarrow\,\langle x_i,\, x_{i+1},\, \dots,\, x_n,\, f(x_i),\, f(x_{i+1}),\, \dots,\, f(x_n) \rangle \in R$.   
\end{defn}    

\begin{defn}[\emph{Source Test Case and Follow-Up Test Case}]
	\label{def:sourcefollowup}
	{\ \\}
	Consider a given MR $R$ with $R(x_i,\, x_{i+1},\, \dots,\, x_n,\, f(x_i),\,\\ f(x_{i+1}),\, \dots,\, f(x_n))$.
	All inputs $x_i$ and outputs $f(x_i)$ for $0 \leq i < n$ are referred to as \emph{source inputs} and \emph{source outputs}, respectively, and the correlating test cases
	$\overline{x_i}$ as \emph{source test cases}. From there, suppose that new test cases $\overline{x_j}$ are created based on the existing source test cases, with 
	$m > j > n$. Then all $x_j$ are referred to as \emph{follow-up inputs} and their test cases $\overline{x_j}$ as \emph{follow-up test cases}.
	A sequence of source and follow-up test cases (or inputs) is called a \emph{Metamorphic Group (MG)}, i.e. $\langle \overline{x_i},\, \overline{x_{i+1}} \rangle$ 
	or $\langle {x_i},\, {x_{i+1}} \rangle$.
\end{defn}

\begin{defn}[\emph{Metamorphic Testing}]
	\label{def:MT}  
	{\ \\}
	Let $R$ be an already identified MR	with $R(x_i,\, x_{i+1},\, \dots,\, x_n,\\ f(x_i),\, f(x_{i+1}),\, \dots,\, f(x_n))$.
	Metamorphic testing (MT) of $P$ based on $R$ involves the following (remaining\footnote{MR identification is also a step within MT, to be precise \cite{founders}.}) steps: \cite{challengesOpportunities}
	\begin{enumerate}
		\item\label{def:MT:step1} Project the found MR onto the implementation, i.e. replace $f$ by $P$ in $R$ and use the inputs as test cases.
		
		\item\label{def:MT:step2} Execute $P$ applying the source test cases $\langle \overline{x_i},\, \overline{x_{i+1}},\, \dots,\\ \overline{x_n}\rangle$ in order to obtain their respective outputs
			$\langle P(\overline{x_i}),\\ P(\overline{x_{i+1}}),\, \dots,\, P(\overline{x_n}) \rangle$.
		
		\item\label{def:MT:step3} Construct and execute a number of follow-up test cases $\langle \overline{x_{j}},\, \overline{x_{j+1}},\, \dots,\, \overline{x_m}\rangle$ in accordance with the relation $R$
		and obtain their respective outputs $\langle P(\overline{x_{j}}),\, P(\overline{x_{j+1}}),\, \dots,\, P(\overline{x_m}) \rangle$. 
		
		\item\label{def:MT:step4} Decide whether $R$ is violated. If $R$ is not satisfied, then $R$ revealed that $P$ contains faults. Formally,
		\\$
		\langle 
			\langle 
				\overline{x_{i}},\, \overline{x_{i+1}},\, \dots,\, \overline{x_n}, P(\overline{x_i}),\, P(\overline{x_{i+1}}),\, \dots,\, P(\overline{x_n}) 
			\rangle,
			\, 
			\\\langle 
				\overline{x_{j}},\, \overline{x_{j+1}},\, \dots,\, \overline{x_m},\, P(\overline{x_{j}}),\, P(\overline{x_{j+1}}),\, \dots,\, P(\overline{x_m}) 
			\rangle
		\rangle\\ 
		\notin\,R\,\Rightarrow\,P\,is\,faulty 
		$
	\end{enumerate}  
\end{defn}

\subsection{Metamorphic Relation Identification}
Previous studies found that due to the pitfalls involved in MR identification, the process constitutes a major challenge in MT \cite{challengesOpportunities}. One commonly misunderstood concept is that, albeit MRs being necessary properties of the target program, not all necessary properties are MRs. Another misconception is that not all MRs are equality relations, although many of them are vastly defined as such. An invaluable skill within the process of MR identification is the ability to distinguish good MRs from less effective ones, as deducible from several studies \cite{survey, selection_good_MRs, case_study_useful_mrs}. For instance, Chen et al. \cite{case_study_useful_mrs} emphasised that MRs should make the execution of follow-up test cases as different as possible from that of the source test cases, which has been confirmed by several later studies. They also reported that knowledge of the problem domain and the program structure is required for the determination of good MRs, whereas purely theoretical knowledge of the problem does not suffice. The survey further concludes that MRs ought to be formally described, as for instance pointed out by Hui and Huang \cite{hui_mr_decomp}.
Another important property of MRs is diversity (i.e. they should involve distinct parameters) as suggested by Liu et al. \cite{howEffectively}. Mayer and Guderlei \cite{selection_good_MRs} have also conducted a study on the selection of good MRs and define good MRs to be rich in semantics of the system under test, and that being too close to the implementation limits the usefulness of MRs (e.g. when the difference between the source and follow-up test cases is balanced out by the implementation) \cite{survey,selection_good_MRs}.

In the following we present two of the MRs that we defined for this project. Further, we add the following notation for specific parameters concerning an output $f(x_i)$:\\    
$f(x_i)_d$\,\; the distance travelled by the drone(s) in metres,\\
$f(x_i)_t$\,\,\; the wall clock time passed in seconds,\\
$f(x_i)_{ac}$\, the number of performed avoidance manoeuvres,\\
$\Delta_d, \Delta_t$\,\, tolerance values $\geq 0$ for distance and wall clock time;
{\ \\}
\subsubsection{Source Inputs and Follow-up Inputs}
{\ \\}
\vspace{-5mm}
\begin{table}[h]
\resizebox{\columnwidth}{!}{%
\begin{tabular}{|l|l|}
\hline
\textbf{Source Input}    & \textbf{Follow-up Input}
\\ \hline
\begin{tabular}[c]{@{}l@{}}$x_1:$ obstacles, one drone, \\ path A $\rightarrow$ B\end{tabular}                                                    & \begin{tabular}[c]{@{}l@{}}$x_2:$ obstacles, one drone, \\ path B $\rightarrow$ A\end{tabular}                                                    \\ \hline
\begin{tabular}[c]{@{}l@{}}$x_3:$ no obstacles, two drones, \\ separate paths A$_1$ $\rightarrow$ B$_1$,\\ A$_2$ $\rightarrow$ B$_2$\end{tabular} & \begin{tabular}[c]{@{}l@{}}$x_4:$ no obstacles, two drones, \\ separate paths B$_1$ $\rightarrow$ A$_1$,\\ B$_2$ $\rightarrow$ A$_2$\end{tabular} \\ \hline
\end{tabular}
}
\end{table}
\subsubsection{Metamorphic Relations}
\label{subsub:MR}
{\ \\}
$R_1: \langle x_1,\, x_2,\, f(x_1),\, f(x_2) \rangle \in R_1 \iff f(x_1)_{ac} \geq 1 \,\wedge\\ f(x_2)_{ac} \geq 1  $.\\
This relation is aimed at testing the overall reliability of object avoidance using stationary obstacles. The number of required avoidance manoeuvres might differ due to an asymmetrical shape of the obstacle and/or the locations of A and B.
The constraint that the number of avoidance manoeuvres should be at least one for each path has therefore been
chosen and shall ensure that avoidance is active and working.\\
$R_2: \langle x_3,\, x_4,\, f(x_3),\, f(x_4) \rangle \in R_2 \iff $for each drone$:
 f(x_3)_{ac} = f(x_4)_{ac} = 0 \,\wedge\, f(x_3)_d = f(x_4)_d \pm \Delta_d \,\wedge\, f(x_3)_t = f(x_4)_t \pm \Delta_t$.\\
This is one of the relations that involves two drones simultaneously. The object influence threshold and drone positions are chosen in a way such that the drones are separated just widely enough not to trigger object avoidance. Hence, we demand that the avoidance manoeuvre count should stay zero as object avoidance would soon be triggered once a drone wanders from its shortest path. Moreover, the travelled distances and taken durations should also be approximately identical for each drone among both paths.

\section{Implementation}\label{sec:implementation}
In this section, we describe the system architecture, tools and frameworks, and APIs used as well as the parameters used within the simulator to run our simulation. 
Since our aim is to investigate the applicability of metamorphic testing in autonomous systems using simulations, we demonstrate the testing approach by implementing an object avoidance task, which is a non-trivial exercise in ASs.


\subsection{System Setup}
The technical setup consists of the simulation environment with the autonomous drone and required plugins, the ArduPilot software along with MAVProxy and the software in the loop (SITL) simulation,    
and the robot operating system (ROS) with the MAVROS package, using the MAVLink (Micro Aerial Vehicle Link) communication protocol \cite{obstacle_avoidance_lidar}. For programmatically controlling the drone as well as performing obstacle detection and avoidance, the DroneKit \cite{dronekit_python} and GNC (Guidance Navigation and Control) \cite{iq_gnc} APIs are used. The entire simulation and implementation has been conducted on a local system running Ubuntu 20.04.2 LTS 64-bit.

\subsubsection{Gazebo}
One of the most popular simulators in robotics which was used to implement the simulation environment for this project.
In addition to its popularity, we considered its superior online documentation and community support in our decision to opt for Gazebo \cite{sim_popularity_comparison}. Common alternatives are Unity \cite{comparison_gazebo_vs_unity}, Webots, Simbad, or MRDS (Microsoft Robotics Developer Studio) \cite{simulators_survey}. We use given open-source models from Open Robotics\footnote{Available online at \url{http://models.gazebosim.org} [accessed Jan-2022].}. Green landing platforms are used as starting points, while red landing platforms act as destination points.  

\subsubsection{ArduPilot SITL}
ArduPilot is an open-source autopilot framework which supports a wide variety of hardware such as various aircraft, ground vehicles, or underwater vehicles \cite{ardupilot}. A popular alternative is PX4 autopilot, while 
ArduPilot subjectively seems to be tested more thoroughly and to be more reliable \cite{iq_sim}. The ArduPilot SITL (Software In The Loop) simulation provides a virtual way of using the vehicles, allowing developers to test their behaviour realistically without hardware.

\subsubsection{MAVProxy}
A minimalistic ground control station for MAVLink\footnote{A messaging protocol for communicating with(in) drones.}-based systems. It allows the operator to control the drone through setting parameters or directly sending commands. It is predominantly used by developers as it is command-line based and not very intuitive to use, however it can be complemented with another ground station like QGroundControl or Mission Planner, which come with a GUI (Graphical User Interface). The MAVProxy console is launched along with SITL which is able to read the messages
sent from MAVProxy.

\subsubsection{Robot Operating System (ROS)}
An open-source robot operating system proposed by Quigley et al. in 2009 \cite{ros_founders} which offers a communication layer above the host OS. It comes with a set of very useful tools for developing and programming robots. For this project, ROS Noetic was used along with the Catkin build environment for ROS packages. Essential ROS concepts are nodes, messages, topics, and packages (e.g. MAVROS).
\subsection{Guidance Navigation and Control (GNC) API}
The GNC API \cite{iq_gnc} from Intelligent Quads Simulations \cite{iq_sim} provides a software collection designed for drone developers with a large amount of documentation and an active community. It implements object detection and avoidance using an interesting principle known as the \emph{Potential Field Method} \cite{obst_avoidance_algorithm}. This concept was first proposed by O. Khatib in 1985 \cite{concept_for_potential_fields} and suggests creating a potential field where the obstacles generate repulsive potential similar to an energy field. The robot is considered as a particle immersed 
in this field desiring to reach a destination, while the repulsive force prevents it from getting close to the obstacles \cite{obst_avoidance_algorithm}. This repulsive potential has an increasing influence when the robot moves closer to the obstacle and decreases upon gaining distance from the obstacle. Let $d_{obst_i}(p)$ be the minimal distance from the robot's position $p := $ \verb+[X, Y, Z, +$\psi$\footnote{The orientation of the drone in degrees.}\verb+]+ to the $i$'th obstacle $obst_i$, let $k_{obst_i}$ be a scaling 
constant and $d_0$ be the object influence threshold\footnote{The range from which avoidance is triggered.}. Then the repulsive potential generated by obstacle $i$ 
dependent on the robot's position $p$ can be defined as:  
\begin{equation} \label{eq:repulsive_potential}
	U_{rep_i}(p) = 
	\begin{cases}
		\frac{1}{2}k_{obst_i} \Big( \frac{1}{d_{obst_i}(p)} - \frac{1}{d_0} \Big) ^2 \;\;\;\text{if } d_{obst_i}(p) < d_0 \\
		0 \;\;\;\;\;\;\;\;\;\;\;\;\;\;\;\;\;\;\;\;\;\;\;\;\;\;\;\;\;\;\;\;\;\;\,\,\,\, \text{if } d_{obst_i}(p) \geq d_0
	\end{cases} 
\end{equation}

The API uses a two-dimensional $360$-degree Hokuyo Lidar (Light Detection and Ranging) \cite{obstacle_avoidance_lidar} sensor for recording the distance data which are being published to a ROS topic. The sensor contains numerous laser rays each of which is able to sense nearby objects and log the data as to where the laser ray is interrupted. The data are then retrieved and parsed by the API in order to perform the aforementioned avoidance algorithm.  

\subsection{Testing}
\begin{figure}[h]
	\centering
	\includegraphics[width=\linewidth]{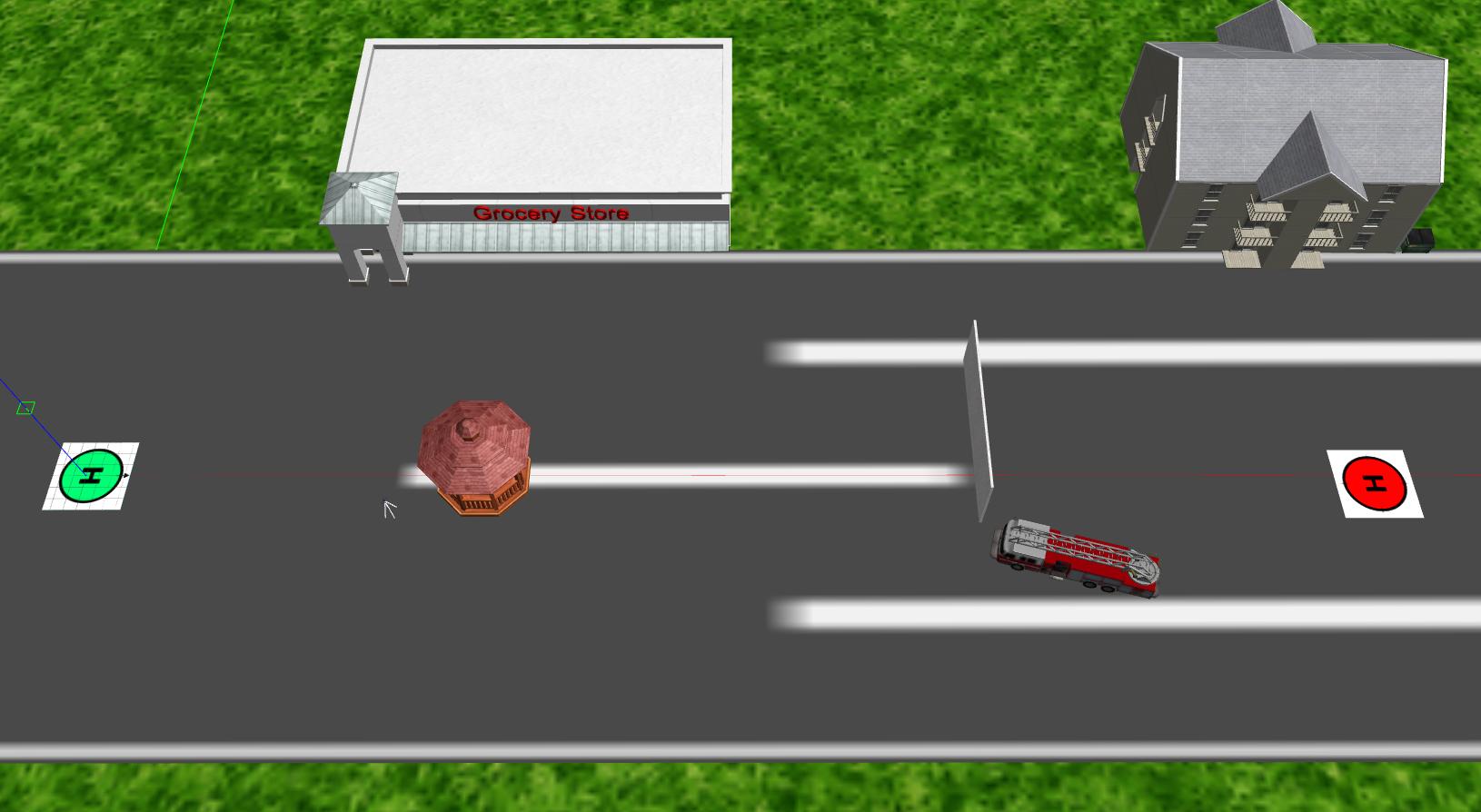}
	\caption{$R_1$: One drone moving between two targets, avoiding obstacles.}
	\label{fig:R1_sim}
\end{figure}

Several functionalities had to be implemented in order to apply the MRs defined in section \ref{subsub:MR}. For instance, functions to measure the travelled distance and passed time, to monitor avoidance behaviour and count the number of avoidance manoeuvres, as well as appropriate configurations to control multiple drones models simultaneously without conflicts. We conducted MT using four carefully defined MRs as well as several alterations of them along with further conventional testing. All test cases have been run multiple times so as to ensure consistency. We generally used an object influence range from $0.35$ to $10$ metres, a velocity of $1$ metre/second, takeoff height of $2$ metres and a ``waypoint-reached-tolerance'' of $0.5$ metres (varying parameters have been used for some further testing). Both the distance measurement and avoidance detection mechanisms update once per second, which is accurate enough while only causing minor performance overhead. The launch points and destination targets were positioned $100$ metres apart from each other such that the results across various relations are comparable. Test cases included those with no obstacles at all as well as those with obstacles placed in the path connecting the targets. 
Figure \ref{fig:R1_sim} shows a screenshot of the Gazebo world during a test run of $R_1$.  

\section{Results and Discussion}\label{sec:results}
Within this section we present the findings that were revealed during MT using our MRs. We offer a brief interpretation of these results and also demonstrate a few limitations. We also evaluate the suitability of MT as a testing strategy for researchers and software engineers in the AS domain.

\subsection{Consistency without Avoidance}
Our first goal was to ensure that the drones themselves show comparable behaviour across equivalent
situations without yet paying regard to object avoidance. Fortunately, the relations which did not assess object avoidance (e.g. $R_2$) showed highly consistent results. Even for varying parameters (e.g. drone velocity, object influence threshold etc.) the resulting distances and durations stayed close together and the relations validated successfully almost all the time. There has been a single occurrence of a false validation of $R_2$, where reviewing the logs showed that one drone took a few seconds to recognise a reached waypoint and move on, causing the durations to differ by approximately $7$ seconds.
Whenever this delay exceeds a certain amount, the tolerated discrepancy $\Delta_t$ will be too strict, leading to a false validation which is precisely the intention of this relation. Since the cause for the delay seemed to be simple overload, we adjusted the previously used parameter $\Delta_t = 3$s to $10$s, with $\Delta_d = 1$m and have not encountered any failed validations henceforth. The relation $R_2$ further demonstrated that the object influence threshold is working as intended since the drones were placed approximately $25$ metres apart using an object influence threshold of $10$ metres, and further tests have been conducted using a threshold of $12$ metres. This means that there was only $1$ metre of tolerance between the drones along their entire path, though we still have not registered any avoidance manoeuvres. Thus, our testing proved that the drone behaviour and path selection is fairly consistent when leaving object avoidance out of consideration, indicating that testing with object avoidance is reasonable. 
\subsection{Challenges of Object Avoidance}
During our testing with the MRs that regarded object avoidance, several interesting properties as well as some limitations of the implementation were revealed.

\subsubsection{Object Influence Threshold}
The region around the drone where object avoidance is triggered played a decisive role during our testing in regards to object avoidance: The object influence threshold ought to be large enough according to the set drone velocity, such that the drone has enough space to brake before an obstacle. However, this threshold is static and therefore also viable during avoidance, where the velocity is very low. This causes the drone to slowly distance itself from the obstacle until the 
threshold is reached, move closer again and repeat the process, resulting in a zigzag-shaped path around the obstacle and multiple detected avoidance manoeuvres. For this reason the efficiency of avoidance manoeuvres is severely compromised and the potential field method may not be suitable for robots that are destined to reach a target as quickly as possible. This avoidance issue was revealed by all MRs that assessed object avoidance and e.g. becomes evident immediately from the data logs during the test runs. For instance, the output shown by Figure \ref{fig:R1_output} yields that the number of avoidance manoeuvres is higher than that of obstacles (as seen in Figure \ref{fig:R1_sim}).
\begin{figure}[h]
    \centering
    \begin{BVerbatim}[fontsize=\footnotesize]
...
Waypoint 1 has been reached!
Shortest path to Waypoint:         100.196 m
Distance travelled:                189.520 m
Elapsed wall clock time:           251.815 s
Number of avoidance manoeuvres:     12
...
Waypoint 2 has been reached!
Shortest path to Waypoint:         100.534 m
Distance travelled:                126.710 m
Elapsed wall clock time:            76.002 s
Number of avoidance manoeuvres:      6
...
---------- MR Validation ----------
MR 1: True
    \end{BVerbatim}
    \caption{Console output snippet of a test run using $R_1$.}
    \label{fig:R1_output}
\end{figure}

\subsubsection{Number of Avoidance Manoeuvres}
\begin{figure}[h]
    \centering
    \vspace{-9mm}
    \resizebox{\columnwidth}{!}{\input{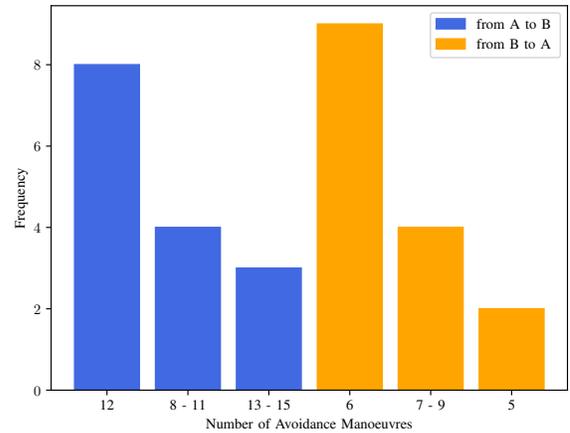}}
    \caption[Number of avoidance manoeuvres]{The most frequent avoidance manoeuvre counts using $R_1$.}
    \label{fig:avoidance_maneuvers}
    \vspace{-3mm}
\end{figure}

The previous figure further illustrates that the drone performed twice as many avoidance manoeuvres on one direction compared to the other one, meaning that different paths have been taken by the drone. The results of $R_1$ showed that the drone's path A $\rightarrow$ B is generally different from and less optimal than the reverse path B $\rightarrow$ A as in that it requires more avoidance manoeuvres (and therefore more time). The most common avoidance manoeuvre counts were roughly $12$ for the path A $\rightarrow$ B and $6$ for the path B $\rightarrow$ A, which is illustrated in Figure \ref{fig:avoidance_maneuvers} on the basis of $15$ test executions using $R_1$. The parameters time, distance, and avoidance manoeuvres generally showed to correspond to each other, as in that a large number of avoidance manoeuvres indicated more time passed and distance travelled, and vice versa.  

\subsubsection{Moving Obstacles}
\begin{figure}[h]
    \centering
    \resizebox{\columnwidth}{!}{\input{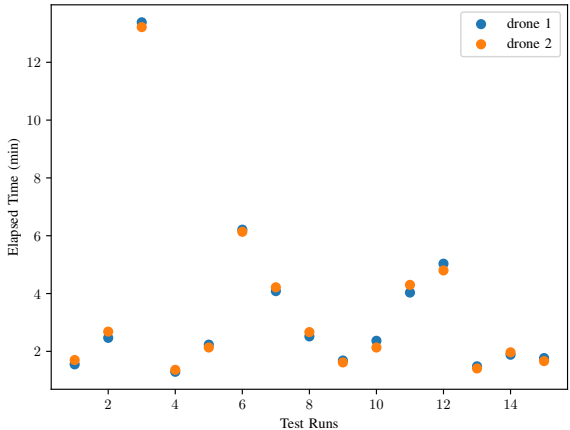}}
    \caption{The durations taken by each drone to perform a mission with intersecting paths.}
    \vspace{-3mm}
    \label{fig:durations}
\end{figure}

One of our relations investigated the behaviour of two drones launching from separate starting points, though destined to the same target. Thus, an avoidance manoeuvre by both drones was expected given the drones and their flight parameters are identical. In spite of the drones never crashing and proving consistency of the Lidar sensor, there have been occasions of rather unfortunate avoidance behaviour. The drones would keep symmetrically avoiding each other
towards the same direction and continue to move further and further away from the target. Consequently, one test run took over $13$ minutes for both drones to complete their paths, while it took between one and three minutes in most other test executions. The total time taken by each drone to perform its mission is shown by Figure \ref{fig:durations} based on $15$ example test runs. In spite of the highly inconsistent observations as well as the overall fluctuating durations, it was detected that
the durations required by each drone stayed highly similar within a single test run. 
The occasional deviations mostly traced back to one drone taking a tiny amount of time longer to take off or to detect its waypoint, or ending up ahead/behind the other drone due to an
asymmetrical avoidance manoeuvre, which then has a major influence on the subsequent behaviour.
Overall, it was surprising to see how random the drones behave albeit being controlled exactly the same way.
\vspace{-8mm}

\subsubsection{Distance Vector Conflicts}
We have observed a bug where a drone would avoid an obstacle and then suddenly start registering to have reached waypoints and ultimately land, despite certainly not being close to any waypoint. The function that checks whether a waypoint has been reached seemed to work just fine upon investigation and adjusting the control loop rate seemed to partly alleviate the issue, surprisingly.
We have discussed this bug with one of the API's developers, Sahasrajit A., who recommended suspending waypoint pursuit whenever object avoidance is active. This is because the drone might be confused as to where to go during avoidance due to conflicting distance vectors (the avoidance algorithm demands to move away from the obstacle whereas the waypoint pursuit instruction concurrently orders the drone to move forward). With the appropriate implementation changes in place, we have been unable to reproduce this issue henceforth.

\subsubsection{Limitations of the Sensor}
The relation $R_1$ further demonstrated that certain, complex shapes of obstacles may pose a challenge for the two-dimensional sensor. Consider a fire truck model (visible in Figure \ref{fig:R1_sim}) which has a folded aerial ladder mounted on its roof, increasing the vehicle's height at its centre. When the drone is travelling at the height of the truck's ladder, it is possible that the drone moves too close and slightly touches the truck with its standoffs. Such an incident occurred twice and essentially traces back to the two-dimensional nature of the sensor as well as the obvious fact that the drone model is larger in height than the line length of the laser rays\footnote{\href{https://www.vision-doctor.com/en/illumination-calculations/calculation-of-laser-line-length.html}{Laser Line Length} [accessed Apr-2022]}. Increasing the object influence threshold will alleviate this issue, while altering the sensor would probably be a better solution. Further testing has again brought up the two-dimensionality of the sensor in the case of higher wall heights. The most efficient way of avoiding this wall would clearly be to increase the flight height  where the wall is no longer sensed and continue to move forward, avoiding the wall from above. However, the sensor is unaware of the wall having only insignificant height and is going to demand the drone to move left or right until the wall ends. Metamorphically, this difficulty can be revealed through an extensive amount of time taken and/or distance travelled in spite of a small distance of the shortest path. The purely horizontal behaviour is not only inefficient in this case, but may also lead to a complete failure in a situation where the drone is completely surrounded by obstacles. It becomes more and more evident that a single, two-dimensional sensor is probably not sufficient for a real-world application and vertical avoidance capabilities would be desirable. It was also found that the number of avoidance manoeuvres may be useful for evaluating path efficiency: The less avoidance manoeuvres are performed, the more linear the path is going to be which equals a higher efficiency. This dependency could be utilised for a Machine Learning approach where the drone learns to choose more efficient paths. An AI drone could sense that the path B $\rightarrow$ A required notably fewer manoeuvres while connecting the same points A and B, and hence choose this path over the path A $\rightarrow$ B in the future. Besides, the fact that the most frequent avoidance manoeuvre counts were far apart from each other across the paths A $\rightarrow$ B and B $\rightarrow$ A shows that obstacle positions along with the drone's launch point heavily influence the path chosen by the drone. The results of using other drones as moving obstacles underlined that the autonomous system in combination with the used algorithm is not stable as in that it does not behave essentially the same way for exactly the same situation. In fact, it would require intelligent path selection or waiting behaviour in order to attain good efficiency in such missions. This could be achieved by an approach where drones communicate with each other and adjust their paths according to the other drone's destination and position. 
For multiple reasons, vertical avoidance and compatibility with all object shapes would be highly desirable, hence we would recommend inducing a three-dimensional sensor. With such a sensor in place, one could conduct a wide range of further tests using a vertical obstacle course. 

\subsection{Expedience of Metamorphic Testing}
Applying MT to this domain has significantly improved our domain knowledge and brought up several important aspects about the system under test. Even during the process of MR identification, the potential of this testing technique in AS has already started to show: during the construction of $R_1$, it was initially planned to demand that,
\vspace{-1.5mm}
\begin{equation} \label{exp:MR}
f(x_1)_{ac} = f(x_2)_{ac}
\end{equation}

However, the nature of the avoidance algorithm suggests that the drone may not choose an identical path for both path directions, resulting in potentially varying avoidance manoeuvre counts despite fundamentally correct behaviour. This aligns with the reasoning for a longer travel distance expected by the unmanned aerial vehicle (UAV) example for the transformation IV rule presented by Li et al. \cite{MT_UAV}. Thus, formulating a weak MR actually proved to be useful upon realising its deficiencies. This demonstrates how extensively MRs ought to be thought through and how much there is to be gained from concerning oneself with MT in this domain.

\section{Related Work}\label{sec:related} 
Ever since the proposition of MT, researchers and engineers have been applying the technique to real-life applications across different fields such as compilers \cite{compilerValidation}, search engines \cite{searchEngines}, computer graphics \cite{gpuCompiler}, machine learning \cite{AImachineLearning}, autonomous systems \cite{AIdrones, objectDetection} and numerical analysis \cite{MTexamples}. A major survey in MT was conducted by Segura et al. in 2016 \cite{survey} and updated in 2020 \cite{mainArticle}. Segura et al. studied the trends in MT and their results indicate that MT is a thriving topic with an increasing amount of contributions and concludes that it is an effective testing technique for alleviating the oracle problem. When one analyses 
Figure \ref{fig:applications} and compares it to the older survey of 2016, one observes that applications in simulation and modelling have decreased by 4\%, whereas applications in bioinformatics have increased by 3\%. There has also been a 2\% increase in the autonomous vehicles domain even though it was not previously obtainable in the former survey \cite{survey}
at all. This leads to the assumption that the number of applications in this domain used to be negligibly small until 2016, but has been rising in the subsequent years. This has partly motivated out work since there has been a very limited amount of studies on MT in autonomous systems and their simulation hitherto.


Segura et al. \cite{mainArticle} claim that it is essential to comprehend that applications among different domains may ask for structural adaptation within the concept, yielding different approaches to MT. For instance, researchers in compiler testing proposed that removing dead code should not influence the functionality of the compiler-generated code. Their idea is to simply run an initial test case, remove some dead code and re-run the program (“next” test case) treating any observed changes as a failure. This approach has been applied in validating the GCC (GNU Compiler Collection) and LLVM (Low-Level Virtual Machine) \cite{compilerValidation}. Another survey followed in 2018 by Chen et al., which highlights some vital challenges and unveils the most promising opportunities for future research on MT \cite{challengesOpportunities}. The authors recommended for researchers who develop software engineering methods that may require a test oracle to consider MT during development, since this may alleviate the oracle requirement. Liu et al. \cite{howEffectively} provided answers to the question of how effectively MT alleviates the oracle problem, conducting an empirical analysis where MT was performed by different testers and on multiple programs. They conclude that MT does generally alleviate the oracle problem, and argue that the diversity of MRs is much more important than their quantity.

\begin{figure}[h]
	\centering
		\includegraphics[width=0.40\textwidth]{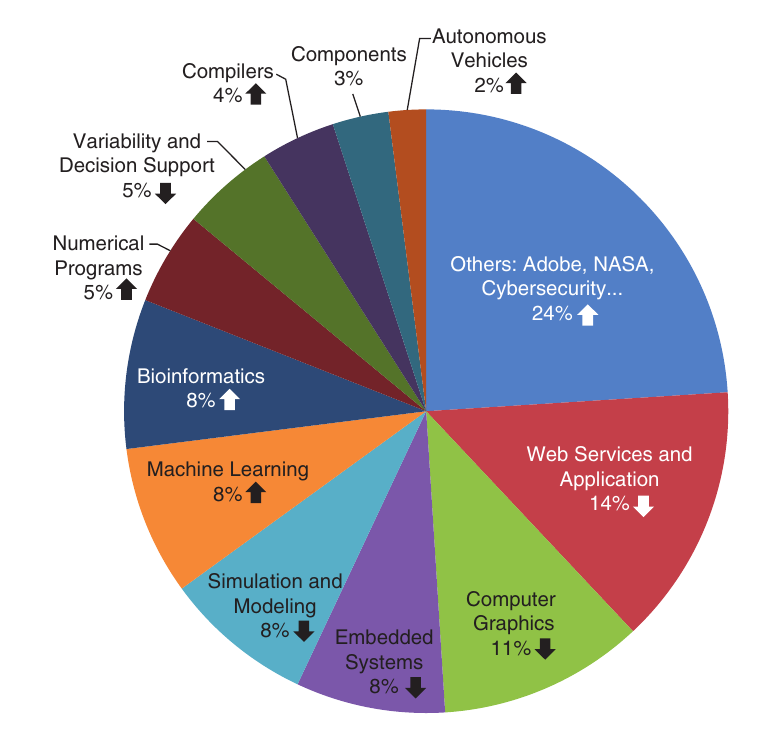}
		\label{fig:applications}
    \vspace{-3mm}
	\caption{MT application domains (2020) \cite{mainArticle}}
    \vspace{-5mm}
	\label{fig:applications}
\end{figure}

In autonomous system simulations, Lindvall et al. \cite{AIdrones} have presented a model-based testing approach combined with MT principles for testing of autonomous drones. In fact, their work posed a major motivation for our project. The authors presented a number of equivalences for their AI drone which represent an outstanding base for finding MRs and constructing test cases. The simulation environment for their project was implemented using Unity 5.
Their approach showed potential for identifying corner cases and thereby increased their confidence in the autonomous system. They proposed to apply the testing to more MRs, use other drones as moving obstacles as well as to add physical conditions such as wind.

In conclusion, the impressive amount of recent contributions to MT suggests that the testing strategy is being recognised as a promising approach for alleviating the oracle problem, although there are only few papers to be found concerning MT in autonomous system simulations. Still, the few applications in autonomous systems showed impressive results which indicate that MT may indeed be an adequate technique for this domain \cite{AIdrones, objectDetection}.
\vspace{-1.5mm}
\section{Conclusion}\label{sec:conclusion}
We have provided an overview of MT as well as an implementation of this testing technique in autonomous drone simulations. In adherence with specific properties of effective MRs, multiple relations have been carefully defined and utilised for testing.  We have demonstrated what Lindvall et al. \cite{AIdrones} proposed in their future work section, i.e. to apply multiple MRs and perform MT in a simulation with multiple drones. We posit that despite a seemingly simple algorithm, complex and diverse observations of drone behaviour may occur. We have also demonstrated a dependency between drone velocity and the object influence threshold, which is a crucial property when utilising the potential field method and causes an unfortunate avoidance behaviour in many cases. 



To conclude, through conducting MT in our simulation we have been able to deduce several essential properties of the system under test that we were previously unfamiliar with. With this, we have investigated some issues that only a few previous studies have dealt with. Each of the implemented metamorphic relations has revealed vital details about the flaws and behaviours of the used software and object avoidance strategy.
Despite the algorithm technically working as intended (for particles), there may be other choices that are more useful to our setting. Due to the complexity of this domain and the typically limited quality of documentation, we trust that this paper and the technical insight it provides may be helpful to other developers. Concerning ourselves with MT in this domain has certainly improved our own understanding of the testing strategy as well as the domain. Both the testing process and the results indicated great applicability for our setting which is why we wish to recommend MT to developers in autonomous systems. In particular, we believe that invoking MT as a complementary addition to conventional software testing is going to have a very positive influence in nearly all cases.

In the future, for the potential field method, we plan to invoke dynamic object influence thresholds according to the drone velocity, such that the threshold increases when the drone accelerates and decreases when it brakes. As a consequence, the efficiency of an avoidance manoeuvre could be massively improved and the observed zigzag-shaped avoidance path could be eliminated. We would also like to incorporate intelligent path planning (selection) by trying an algorithm of a different nature like the Bug algorithms \cite{obst_avoidance_algorithm}, where avoidance paths are generated around obstacles. AI drones could learn to choose the most efficient path according to the amount of required avoidance manoeuvres, whereas for other drones as obstacles they could simply communicate with each other and adjust their paths according to the other drone’s destination and position, similar to what Chen et al. \cite{Chen2020-nd} have done. One could then utilise shape matching algorithms to evaluate the quality of chosen paths and use this for MRs, as Lindvall et al. \cite{AIdrones} have done similarly. 

In the same degree, we plan to implement a 3D obstacle detection method that enables the detection of obstacles all around a test drone for instance, thereby ensuring that obstacle avoidance manoeuvering can happen in a more realistic manner as would be obtainable when the AS is in operation. With a reliable avoidance behaviour and our functionality of detecting a manoeuvre, testers could start to become stricter with the expected number of avoidance manoeuvres in MRs and thus really narrow down the inconsistencies of the used avoidance algorithm.\\

\vspace{-2mm}
\balance
\bibliographystyle{unsrt} 
\bibliography{references}

\end{document}